# BEAM-BEAM INTERACTION IN NOVEL, VERY HIGH LUMINOSITY PARAMETER REGIMES

M. Zobov, INFN LNF, Frascati, Italy

*Abstract*

In order to achieve luminosities significantly higher than in existing machines, future storage-ring based colliders will need to operate in novel parameter regimes combining ultra-low emittance, large Piwinski angle and high bunch charge; implementation of techniques such as a "crab waist" will add further challenges. Understanding the beam-beam interaction in these situations will be essential for the design of future very high luminosity colliders. Recent developments in modelling tools for studying beam-beam effects, capable of investigating the relevant regimes, will be discussed and examples, including tests with crab waist collisions in DAΦNE, will be presented.

## INTRODUCTION

Pushing the luminosity of storage-ring colliders to unprecedented levels opens up unique opportunities for precision measurements of rare decay modes and extremely small cross sections, which are sensitive to new physics beyond the Standard Model.

Present generation lepton factories have been very successful in achieving their design luminosity performances [1]. In these high luminosity colliders with conventional schemes the key requirements to increase the luminosity are: very small vertical beta function $\beta_y^*$ at the interaction point (IP) comparable to the bunch length $\sigma_z$, high intensity multibunch beams colliding with a small crossing angle $\theta$ (and a small Piwinski angle $\Phi=(\sigma_z/\sigma_x)tg(\theta/2)$), large horizontal emittance $\varepsilon_x$ and beam size $\sigma_x$. However, a further substantial luminosity increase, requested by the physics experiments, is hardly possible exploiting the standard collision scheme due to limitations imposed by both beam dynamics of colliding beams and technological challenges.

During past years several novel collision concepts and new collision schemes were proposed in order to overcome the limitations and recently some of these ideas, such as round beam collisions in VEPP2000 [2], crab crossings at KEKB[3] and crab waist collisions at DAΦNE [4], have been tested experimentally.

At present the crab waist collision scheme [5, 6] is considered to be most prominent for the next generation factories since it holds the promise of increasing the luminosity of the storage-ring colliders by 1-2 orders of magnitude beyond the current state-of-art, without any significant increase in beam current and without reducing the bunch length. Indeed, the successful test of crab waist (CW) collisions at DAΦNE and advantages of the CW collision scheme have triggered several collider projects exploiting its potential. In particular, physics and accelerator communities are discussing new projects of a SuperB-factory [7, 8] and a Super-Tau-Charm factory [9] with luminosities about two orders of magnitude beyond those achieved at the present B- and Tau-Charm factories.

During the last two years comprehensive experimental and numerical studies have been carried out in order to:

- Prove that the crab waist concept works as predicted by theory and numerical simulations
- Benchmark numerical codes versus available experimental measurements at DAΦNE.

These tasks are mandatory to be confident in the expected results while designing the future colliders with unprecedented luminosities.

In the first section of this paper we discuss the performances of the present generation lepton factories and intrinsic beam limitations of the conventional collision scheme. In the following section we introduce the crab waist concept with a discussion of its advantages from beam dynamics point of view. Then, we will briefly describe the main experimental results of the crab waist collision test at DAΦNE. Finally, we discuss numerical codes and their modifications necessary to simulate correctly beam-beam effects in the crab waist collisions and compare the DAΦNE experimental data with the numerical simulations.

## PRESENT LEPTON FACTORIES

Present generation lepton factories have been very successful. For example, both B-factories, KEK in Japan and PEPII in USA, have largely exceeded their luminosity design goals. The Italian Φ-factory DAΦNE has exceeded the phase I design luminosity and obtained a luminosity increase by a factor 3 after implementation of a novel crab waist collision scheme. The recently commissioned Tau-Charm factory in Beijing is well advanced on the way to obtain its design luminosity. A collection of articles summarizing the lepton collider performances can be found in [1].

All the present high luminosity factories relied, at least at the beginning of their operation, on the standard strategy of choosing beam parameters to achieve high luminosity. According to the well-known expressions for the luminosity $L$ and beam-beam tune shifts $\xi_{x,y}$:

$$L = N_b f_0 \frac{N^2}{4\pi\sigma_x^*\sigma_y^*} = N_b f_0 \frac{\pi\gamma^2 \xi_x \xi_y \varepsilon_x}{r_e^2 \beta_y^*}\left(1+\frac{\sigma_y^*}{\sigma_x^*}\right)^2$$

$$\xi_{x,y} = \frac{Nr_e}{2\pi\gamma}\frac{\beta_{x,y}^*}{\sigma_{x,y}^*\left(\sigma_x^*+\sigma_y^*\right)}$$

the key requirements for the luminosity increase in a collider at a given energy are the following:
- Higher number of particles per bunch $N$
- More colliding bunches $N_b$
- Larger beam emittance
- Smaller beta functions at the interaction point (IP)
- Round beams $\sigma_x^* = \sigma_y^*$
- Higher tune shift parameters.

The present factories have obtained their good luminosity performances trying to fulfill almost all the above conditions as much as possible except that it was chosen to collide flat bunches ($\sigma_x^* \ll \sigma_y^*$) since it is rather difficult to provide a good dynamic aperture for the round beam case with both vertical and horizontal beta functions low at the IP. Besides, in order to eliminate parasitic collisions in multibunch operation a small horizontal crossing angle $\theta$ was necessary. In the factories a relatively small Piwinski angle ($\Phi < 1$) was mandatory to avoid excessive geometric luminosity reduction and to diminish the strength of synchrobetatron resonances arising from beam-beam interaction with the crossing angle.

However, a further substantial luminosity increase based on the standard collision scheme is hardly possible due to several limitations imposed by beam dynamics requirements: A) In order to minimize the luminosity reduction due to the hour-glass effect (the dependence of the vertical beam size on the longitudinal position along the crossing region) the vertical beta function at the IP can not be much smaller than the bunch length; B) A drastic bunch length reduction is impossible without incurring into single bunch instabilities: bunch lengthening and microwave instabilities due to the beam interaction with the surrounding vacuum chamber. Besides, too short bunches tend to produce coherent synchrotron radiation (CSR) affecting beam quality and leading to a dramatic increase of the power losses; C) A further multibunch current increase would result in different kinds of coupled bunch beam instabilities, excessive power loss due to interactions with parasitic higher order modes (HOM) and increase of the required wall plug power; D) Higher emittances conflict with stay-clear and dynamic aperture limitations, require again higher currents to exploit the emittance increase for the luminosity enhancement; E) Tune shifts saturate and beam lifetime drops due to a strong nonlinear beam-beam interaction.

In order to overcome these limitations several novel collision concepts and new collision schemes have been proposed. The most know are following: round beam collision preserving an additional integral of motion [10]; crab crossing [11, 12]; collision with large Piwinski angle [13] ("superbunch" in hadron colliders [14, 15]); longitudinal strong RF focusing [16]; collision with travelling waist [17]; crab waist collision [5, 6].

Recently some of these ideas, such as round beam collision [2, 18], crab crossing [3] and crab waist [4] have been experimentally tested. At present the crab waist collision scheme is considered to be most attractive for the next generation factories since it holds the promise of increasing the luminosity of the storage-ring colliders by 1-2 orders of magnitude beyond the current state-of-art. Besides, the scheme has been successfully tested and the results of the experimental tests are in a good agreement with numerical simulations [19].

Contrary to the conventional strategy, the crab waist collision scheme (CW) requires small emittance $\varepsilon_x$, larger Piwinski and crossing angle; there is no need to decrease the bunch length and push beam currents far beyond the values already achieved in the present factories. This scheme can substantially increase collider luminosity since it combines several potentially advantageous ideas: collisions with a large Piwinski angle, micro-beta insertions and suppression of beam-beam resonances using the dedicated ("crab waist") sextupoles.

## CRAB WAIST COLLISION SCHEME

Let us consider two bunches colliding under a horizontal crossing angle $\theta$ (as shown in Fig. 1a). Then, the CW principle can be explained, somewhat artificially, in the three basic steps. The **first one** is large Piwinski angle. For collisions with $\Phi \gg 1$ the luminosity L and the beam-beam tune shifts scale as (see, for example, [20]):

$$L \propto \frac{N\xi_y}{\beta_y^*}; \quad \xi_y \propto \frac{N\sqrt{\beta_y^*/\varepsilon_y}}{\sigma_z \theta}; \quad \xi_x \propto \frac{N}{(\sigma_z \theta)^2}$$

Clearly, in such a case, if it were possible to increase $N$ proportionally to $\sigma_z\theta$, the vertical tune shift $\xi_y$ would remain constant, while the luminosity would grow proportionally to $\sigma_z\theta$. Moreover, the horizontal tune shift would drop as $1/(\sigma_z\theta)$.

Differently from [14, 15], in the crab waist scheme the Piwinski angle is increased by decreasing the horizontal beam size and increasing the crossing angle. In this way we can gain in luminosity as well, and the horizontal tune shift decreases. Moreover, parasitic collisions (PC) become negligible since with higher crossing angle and smaller horizontal beam size the beam separation at the PC is large in terms of $\sigma_x$. But the most important effect is that the length of the overlap area of the colliding bunches is reduced, since it is proportional to $\sigma_x/\theta$ (see Fig. 1).

Then, as the **second step**, the vertical beta function $\beta_y$ can be made comparable to the overlap area size (i.e. much smaller than the bunch length):

$$\beta_y^* \approx \frac{2\sigma_x}{\theta} \cong \frac{\sigma_z}{\Phi} \ll \sigma_z$$

It is worth noting that usually it is assumed that $\xi_y$ (see the expression for $L$ in (1)) always reaches the maximum allowed value, the so called "beam-beam limit". So, reducing $\beta_y$ at the IP gives us several advantages:
- Luminosity increase with the same bunch current.
- Possibility of the bunch current increase (if it is limited by $\xi_y$), thus further increasing the luminosity.
- Suppression of the vertical synchrobetatron resonances [21].
- Reduction of the vertical tune shift with the synchrotron oscillation amplitude [21].

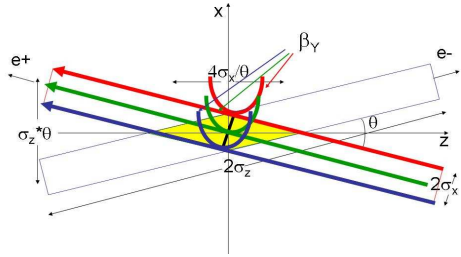

a) Crab sextupoles OFF.

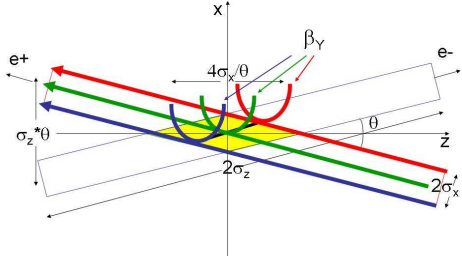

b) Crab sextupoles ON.

Figure 1: Crab Waist collision scheme.

Besides, there are additional advantages in such a collision scheme: there is no need of decreasing the bunch length to increase the luminosity as proposed in standard upgrade plans for B- and Φ-factories. This will certainly help in solving the problems of HOM heating, coherent synchrotron radiation of short bunches, excessive power consumption, etc.

However, implementation of these two steps introduces new beam-beam resonances which may strongly limit the maximum achievable tune shifts. At this point the crab waist transformation enters the game boosting the luminosity. This is the **third step**. As can be seen in Fig. 1b, the beta function waist of one beam is oriented along the central trajectory of the other one. In practice the CW vertical beta function rotation is provided by sextupole magnets placed on both sides of the IP in phase with the IP in the horizontal plane and at $\pi/2$ in the vertical one (as shown in Fig. 2). The crab sextupole strength should satisfy the following condition depending on the crossing angle and the beta functions at the IP and the sextupole locations:

$$K = \frac{1}{\theta} \frac{1}{\beta_y^* \beta_y} \sqrt{\frac{\beta_x^*}{\beta_x}}$$

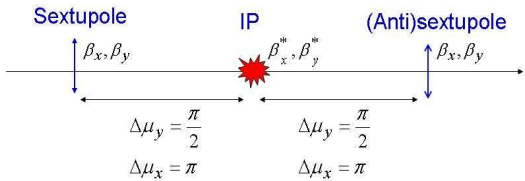

Figure 2: Crab sextupole locations.

The crab waist transformation gives a small geometric luminosity gain due to the vertical beta function redistribution along the overlap area. It is estimated to be of the order of several percent. However, the dominating effect comes from the suppression of betatron (and synchrobetatron) resonances arising (in collisions without CW) from the vertical motion modulation by the horizontal betatron oscillations [22].

Fig.3 demonstrates the resonances suppression applying the frequency map analysis (FMA) for the beam-beam interaction in CW collisions [23]. It shows the beam-beam footprint for DAΦNE with CW sextupoles off (left) and on (right).

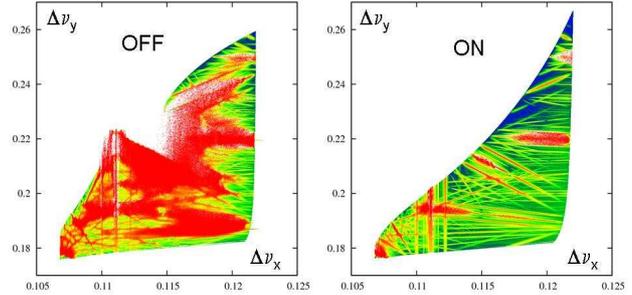

Figure 3: Beam-beam footprint with crab sextupoles off (left) and on (right) obtained by FMA techniques [23].

## EXPERIMENTAL TEST AT DAΦNE

In 2007 the Φ-factory DAΦNE was upgraded implementing the crab waist collision scheme. This required major changes in the design of the mechanical and magnetic layout of both collider interaction regions [24]. Table 1 shows a comparison of the main beam parameters for the DAΦNE upgrade with those of the previous runs for the KLOE and FINUDA experiments.

As one can see from Table 1 the Piwinski angle was increased and the collision region length reduced by doubling the crossing angle, decreasing the horizontal beta function almost by an order of magnitude and slightly decreasing the horizontal emittance. In turn, the vertical beta function at the interaction point was decreased by a factor 2. The crab waist transformation is provided by two electromagnetic sextupoles installed at both ends of the experimental interaction region with the required phase advances between them and the IP. Their integrated gradient is about a factor 5 higher than that of normal sextupoles used for chromaticity correction.

Table 1. DAΦNE best luminosity and respective IP parameters for three experimental runs.

| Parameters | KLOE | FINUDA | Siddharta |
|---|---|---|---|
| Date | Sept 05 | Apr 07 | Apr 09 |
| Luminosity, cm$^{-2}$s$^{-1}$ | 1.53x10$^{32}$ | 1.60x10$^{32}$ | 4.53x10$^{32}$ |
| e– current, A | 1.38 | 1.50 | 1.43 |
| e+ current, A | 1.18 | 1.10 | 1.00 |
| Number of bunches | 111 | 106 | 105 |
| $\varepsilon_x$, mm mrad | 0.34 | 0.34 | 0.25 |
| $\beta_x$, m | 1.5 | 2.0 | 0.25 |
| $\beta_y$, cm | 1.8 | 1.9 | 0.93 |
| Crossing angle, mrad | 2x12.5 | 2x12.5 | 2x25 |
| Tune shift, $\xi_y$ | 0.0245 | 0.0291 | 0.044 |

Right from the start of commissioning, the effectiveness of the new collision scheme was confirmed by several measurements and qualitative observations of the beam-beam behaviour. The simplest and most obvious test consisted in switching off the crab waist sextupoles of one of the colliding beams. This blew up both horizontal and vertical transverse beam sizes of that beam and created non-gaussian tails of the beam distribution, seen on the synchrotron light monitors (Fig. 4). At the same time, a luminosity reduction was recorded by all the luminosity monitors. This behaviour is compatible with the prediction of additional beam-beam resonances when the crab sextupoles are off.

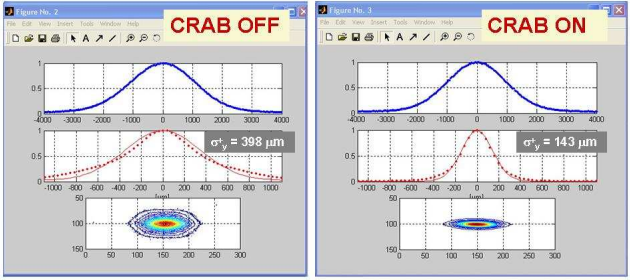

Figure 4: Transverse beam profiles with crab on and off.

The best peak luminosity of $4.53 \times 10^{32}$ cm$^{-2}$s$^{-1}$ was obtained in June 2009 together with a daily integrated luminosity exceeding 15pb$^{-1}$. As one can see from Table 1, the best present luminosity is by a factor 3 higher than that in the runs before the upgrade. The maximum peak luminosity is already very close to the design value of $5 \times 10^{32}$ cm$^{-2}$s$^{-1}$, and work is still in progress to achieve this ultimate goal. The vertical tune shift parameter has been significantly improved and it is now as high as 0.044 (a factor 1.5 higher than before). It is worth mentioning that in weak-strong collisions when the electron beam current is much higher than the positron one the tune shift has reached almost 0.09 (see Fig. 5).

## NUMERICAL ANALYSIS

A comprehensive numerical simulation study has been undertaken for comparison with the experimental data and to prove once more the effectiveness of the crab waist collision scheme. In turn, several dedicated experiments have been carried out at DAΦNE for the numerical codes benchmarking.

To perform the numerical simulations for the DAΦNE upgrade and the future factories based on the CW collision scheme (SuperKEKB, SuperB and SuperC-Tau) we have used the codes that have been successfully used for the present generation e+e- colliders. Both weak-strong (BBC [25], LIFETRAC [26], BBWS) and strong-strong codes (BBSS [27], SBBE [28]) were employed for the beam-beam simulations. However, several modifications have been implemented in order to reproduce better the crab waist collision and/or to save the required CPU time.

The weak-strong codes are very fast (in comparison with the strong-strong ones) and thus are suitable for parameter optimization, luminosity scans, nonlinear beam-beam resonance studies etc. Besides, special CPU saving techniques can be used for beam-beam induced tail simulations and lifetime determination (LIFETRAC).

However, the obvious limitation is that the strong beam remains gaussian while in practice crabbing of the strong beam makes its distribution essentially non-gaussian. Thus, the crab waist transformation can be applied only to the weak beam. In order to overcome this limitation, a new feature was implemented in the LIFETRAC tracking code which allowed calculating the beam-beam kicks from arbitrary strong beam distributions using the prepared-in-advance grid files [29].

A couple of experimental DAΦNE runs were dedicated to tune and to optimize the collider in the weak-strong regime in order to compare measured data with the modified LIFETRAC results. In order to eliminate the crosstalk between e-cloud effects and beam-beam interaction the positron beam was chosen to be the weak one. All the parameters necessary for numerical simulations such as beam currents, transverse beam sizes, bunch length etc. were measured and recorded during these runs. As shown in Fig. 5, practically there is no difference between the numerical predictions and measured luminosity.

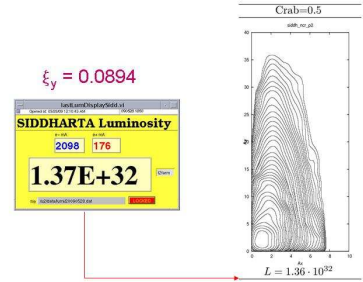

Figure 5: Comparison of measured (left picture - luminosity monitor display) and calculated (right picture) luminosity.

Certainly, the strong-strong simulations can better reproduce collision of high intensity beams since they describe 3D interactions in a fully self-consistent manner. Both beam can be blown up, non-gaussian. The crab waist transformation can be applied to both beams.

The principal limitation in this case consists in a very long CPU time required to describe crab waist collision of tiny beams with the collision area much smaller than the bunch length. A very large number of longitudinal slices are necessary to simulate correctly the beta function variation over this small area.

Among different proposals to solve the problem the following one has been recently realized. In order to reduce the CPU time particle-in-cell simulations are performed only for the central dense area and the Gaussian approximation is used for the tail slices when a separation between the tails becomes more than 5 $\sigma_x$ [30]. Namely, such simulations have demonstrated that a luminosity increase by a factor of 40 is possible in the SuperKEKB factory (see Fig. 6) [31].

Full scale strong-strong simulations for DAΦNE with BBSS and SBBE [32] have shown that the numerical results agree with the measured luminosity within 10%-20%.

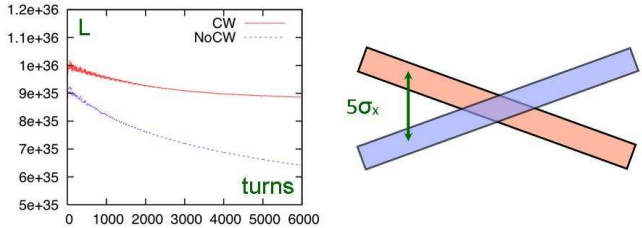

Figure 6: Strong-strong simulations of luminosity for KEKB upgrade (red curve–CW on; green one–CW off).

To complete the CW scheme studies with a kind of control experiment, we have devoted several hours to tuning DAΦNE with the crab sextupoles off. Figure 7 shows a comparison of the luminosity versus the product of the two beam currents obtained with the crab sextupoles on and off. The maximum luminosity reached in the latter case was only $1.6$-$1.7 \times 10^{32}$ cm$^{-2}$s$^{-1}$. Moreover, another drawback becomes very important in collision without the crab sextupoles: in addition to much bigger vertical blow up leading to luminosity decrease, a sharp lifetime reduction was observed at single bunch currents as low as 8-10 mA. For this reason the red curve in Fig. 7 stops at much lower beam currents.

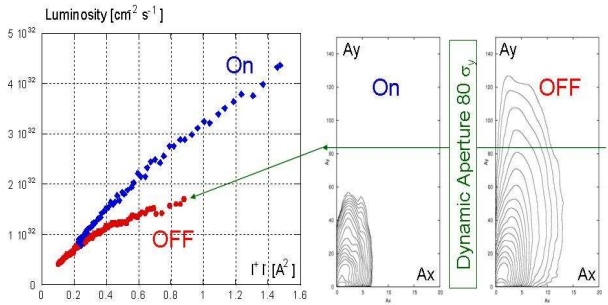

Figure 7: Measured luminosity (left) and simulated beam distribution tails with CW sextupoles on and off.

So, in order to understand the lifetime reduction, we have carried out computer simulations of beam dynamics and luminosity at DAΦNE with a realistic lattice including crab and chromatic sextupoles, damping wigglers, magnet fringe fields etc. During this study the beam-beam interaction simulation was provided by the LIFETRAC code. A tracking of the particles ensemble along the lattice was performed by another software called ACCEL-ERATICUM [33] allowing particles to pass through the variety of storage ring lattice elements in a symplectic way.

The simulations have shown that the crosstalk between the lattice nonlinearities and beam-beam interaction results in vertical beam tail growth (see Fig. 7) and, in the case of the CW sextupoles off, the tails approach the vertical dynamic aperture (estimated to be about 80 $\sigma_y$) at a bunch current as low as 8-10 mA leading to the sharp lifetime reduction that is in agreement with the experimental observations.

## CONCLUSION

The crab waist concept definitely works. This has been proved by the successful test of the novel collision scheme at DAΦNE and by a comparison of numerical simulations with the results of the dedicated experiments.

## ACKNOWLEDGMENTS

The theoretical and numerical studies of beam dynamics in crab waist collision have been done in close collaboration with P.Raimondi, C.Milardi (INFN LNF), D.Shatilov, E.Levichev and P.Piminov (BINP, Novosibirsk), K.Ohmi (KEK, Japan), Y.Zhang (IHEP, Beijing). I am very grateful to the DAΦNE Collaboration Team and Operation Staff for providing the experimental data and the help in performing the dedicated experiments essential for the crab waist collision studies. I also warmly thank M.Preger for his comments on this paper.